\documentstyle[12pt]{article}
\begin{document}
\title{Patterns on  liquid surfaces: cnoidal waves, compactons and scaling}
\author{A. Ludu and J. P. Draayer \\
\normalsize{Department of Physics and Astronomy, Louisiana State
University,} \\
\normalsize{Baton Rouge, LA 70803-4001, U.S.A.}}
\date{\today}
\maketitle
\begin{abstract}

Localized patterns and nonlinear oscillation formation on the bounded free surface of
an ideal incompressible liquid are analytically investigated . Cnoidal modes, solitons
and compactons, as traveling  non-axially symmetric shapes are discused. A
finite-difference differential generalized Korteweg-de Vries equation is shown to
describe the three-dimensional motion of the fluid surface and the limit of long and
shallow channels one reobtains the well known KdV equation. A tentative expansion
formula for the  representation of the general solution of a nonlinear equation, for
given initial condition is introduced on a graphical-algebraic basis. The model is
useful in multilayer fluid dynamics, cluster formation, and nuclear physics since, up
to an overall scale, these systems display liquid free surface behavior.

\end{abstract}


\setlength{\baselineskip} {4ex.}

\vfill
\eject

\section{Introduction}

Liquid oscillations on bounded surfaces have been studied intensively, both theoretically [1-3] and
experimentally [4-6]. The small-amplitude oscillations of incompressible drops maintained by
surface tension are usually characterized by their fundamental linear modes of motion in
terms of spherical harmonics [1-3]. Nonlinear oscillations of a liquid drop introduce new
phenomena and more complicated patterns (higher resonances, solitons, compactons, breakup and
fragmentation, fractal structures, superdeformed shapes) than can be described
by a linear theory.  Nonlinearities in the description of an ideal drop demonstrating
irrotational flow  arise from Bernoulli's equation for the pressure field and from the
kinematic surface boundary conditions [7]. Computer simulations have been carried for non-linear
axial oscillations and they are in very good agreement with experiments [4-6].

The majority of experiments show a rich variety of complicated shapes, many related to the
spinning, breaking, fission and fusion of liquid drops. There are experiments [6] and numerical
simulations [2] where special rotational patterns of circulation emerge:  a running wave originates on
the surface of the drop and then propagates inward.  Recent results (superconductors [8], catalytic
patterns [9], quasi-molecular spectra [10], numerical tests on higher order non-linear equations [11] and
analytical calculations on the non-compact real axis [12-13])  show shape-stable traveling waves for
nonlinear systems with compact geometry.  Recent studies showed that a
similar one-dimensional analysis for the process of cluster emission from heavy nuclei and
quasi-molecular spectra of nuclear molecules yields good agreement with experiment [10]. Such
solutions are stable and express to a good extent the formation and stability of patterns, clusters,
droplets, etc. However, even localised, they have nor compact support neither periodicity (excepting
some intermediate steps of the cnoidal solutions, [10,13]), creating thus difficulties when analysing on
compact surfaces.

In the present paper we comment on the cnoidal-towards-solitons solution investigated in [9,12],
especially from the energy point of view. We introduce here a new nonlinear 3-dimensional dynamical
model of the surface, in compact geometry (pools, droplets, bubbles, shells), inspired by [12], and we
investigate the possibilities to obtain compacton-like solutions for this model. We also study the scale
symmetries of such solutions.

The model in [9,12] consider the nolinear hydrodynamic equations of the surface of a liquid
drop and show their direct connection to KdV or MKdV systems. Traveling solutions that
are cnoidal waves are obtained [10,13] and they generate multiscale patterns ranging from small
harmonic oscillations (linearized model), to nonlinear oscillations, up to solitary waves. These
non-axis-symmetric localized shapes are  described by a KdV Hamiltonian system, too,
which
results as the second order approximation of the general Hamiltonian, next corrextion from the linear
harmonic shape oscillations. Such rotons were observed experimentally when the shape
oscillations of a droplet became nonlinear [4,6,8,13].

\section{Liquid drop cnoidal and soliton solutions from Hamiltonian approach}

The dynamics governing one-dimensional surface oscillations of a perfect ($\rho =$const.), irrotational
fluid drop (or bubble, shell) can be described by the velocity field $\Phi$ and a corresponding
Hamiltonian [1-3,7,10,13].  By expanding the Hamiltonian and dynamical equations in terms of a small
parameter, i.e. the amplitude of the perturbation $\eta $ over the
radius of drop $R_0$, the usual linear theory is recovered in the first order. Higher order non-linear
terms introduce deviations and produce large surface oscillations like cnoidal waves [7].
These oscillations, under conditions of a rigid core of radius $R_0 -h$ and non-zero angular momentum,
transform into solitary waves. In the following, by using the calculation developed in [10],
we present the Hamiltonian approach for the liquid drops nonlinear oscillations. However, this approach
is different from the nuclear liquid drop model point of view in [10], since we do not use here
the nuclear interaction (shell corrections) responsible for the formation of different potential valleys.

The total hydrodynamic energy $E$ consists of the sum of the kinetic $T$ and potential $U$ energies of
the liquid drop. The shape function is assumed to factorize, $r(\theta ,\phi , t)=R_0 (1+g(\theta
)\eta (\phi ,t))$. All terms that depend on $\theta $ are absorbed in the coefficients of some
integrals and the energy reduces to a
functional of $\eta $ only.  The potential energy is given by the surface energy $U_S =\sigma
({\cal A}_{\eta }-{\cal A}_{0})|_{V_{0}}$, where $\sigma $ is the surface pressure coefficient,
${\cal A}_{\eta }$ is the area of the deformed drop, and ${\cal A}_0$ the area of the spherical
drop, of constant volume  $V_0$. 
The kinetic energy 
$T=\rho \oint_{\Sigma }\Phi \nabla \Phi \cdot d{\vec S}/2$, [1-3,10,13],
the kinematic free surface boundary condition $\Phi _r=\partial _t r +(\partial _{\theta }r)
\Phi _{\theta
}/r^2 +(\partial _{\phi }r)\Phi _{\phi }/r^2 \sin \theta $, and the  boundary condition
for the radial velocity  on the inner surface
$\partial _r  \Phi | _{r=R_0 -h}=0$, [7], result in the expression [2,3,10]
\begin{eqnarray}
T={{R_{0}^{2}\rho } \over 2}\int_{0}^{\pi}\int_{0}^{2\pi}
{{
R_0 \Phi \eta _t \sin \theta  +{1 \over {R_0}}
g\eta _{\phi } \Phi \Phi _{\phi }(1-\sin \theta )} \over 
{\sqrt{1+g_{\theta }^{2}\eta ^2 +g^2 \eta _{\phi }^{2}}}} d\theta d\phi .
\end{eqnarray}
If the total energy, written in the second order in $\eta $, is taken to be a
Hamiltonian $ H[\eta ]$, the time derivative of any quantity
$F[\eta ]$ is given by $F_t =[F,H]$. Defining $F=\int_{0}^{2\pi} 
\eta (\phi -Vt)d\phi $  it results ([10], last reference) 
\begin{eqnarray}
{{dF}\over {dt}}=\int_{0}^{2\pi }\eta _t d\phi =\int_{0}^{2\pi}
(2 C_2 \eta_ {\phi }+6 C_3 \eta \eta _{\phi }-2 C_4 
\eta _{\phi \phi \phi }) d\phi =0,
\end{eqnarray}
which leads to the KdV  equation.
Here  
${\cal C}_{2}=\sigma R_{0}^{2}(S_{1,0}^{1,0}+S_{0,1}^{1,0}/2) + 
R_{0}^{6}\rho V^2
C_{2,-1}^{3,-1}/2$, ${\cal C}_{3}=\sigma R_{0}^{2}S_{1,2}^{1,0}/2+
R_{0}^{6}\rho V^2 (2S_{-1,2}^{3,-1}R_0+S_{-2,3}^{5,-2}+ R_0 
S_{-2,3}^{6,-2})/2$,
${\cal C}_{4}=\sigma R_{0}^{2}S_{2,0}^{-1,0}/2$, with
$S_{i,j}^{k,l}=R_{0}^{-l}\int_{0}^{\pi}h^{l}g^{i}g_{\theta }^{j}{\sin 
}^{k}\theta d\theta $.
Terms proportional to $\eta \eta _{\phi }^{2}$ can be neglected
since they introduce a factor $\eta _{0}^{3}/L^2 $ which is small compared
to $\eta _{0}^{3}$, i.e. it is in the third order. In order to verify the correctness of the
above approximations, we present, for a typical soliton
solution  $\eta (\phi ,t)$, some terms occuring in the expresion of $E$, Fig. 1. 
All details of calculation are given in [10,13].
Therefore, the energy of the non-linear liquid drop model can be interpreted as the Hamiltonian of the
one-dimensional KdV equation.
The coefficients in eq.(2) depend on two stationary functions
of  $\theta $ (the depth $h(\theta )$ and the transversal profile $g(\theta )$), hence, under the
integration, they involve only  a parametric dependence.

The KdV equation has the following cnoidal wave (Jacobi elliptic function) as exact solution
\begin{eqnarray}
\eta =\alpha _3 +(\alpha _2 -\alpha _3 )sn^2 \biggl ( \sqrt{{{C_3 (\alpha _3
-\alpha _2 )}\over {12C_4 }}}(\phi -Vt)\bigg | m \biggr ) ,
\end{eqnarray}
where $\alpha _1 ,\alpha _2 ,\alpha _3 $ are constants of integration, $m^2 =(\alpha _3 -\alpha _2
)/(\alpha _3 -\alpha _1 )$. 
This solution  oscillates between $\alpha _2$ and $\alpha _3$,
with a period
$T=2K(m)\sqrt{{(\alpha _3 -\alpha _2 )C_3 }\over {3C_4 }}$, where $K(m)$ is the period of a Jacobi
elliptic function $sn(x | m)$. 
The parameter $V$ is the velocity of the cnoidal waves and  $\alpha _1 +\alpha _2 +\alpha _3
={{3(V-C_2 ) }\over {2C}}$. In the limit
$\alpha _1 =\alpha _2 =0$  the solution eq.(3)
approaches
\begin{eqnarray}
\eta =\eta _0 sech ^2 \biggl [
\sqrt{{{\eta _0 C_3 }\over {12 C_4}}}(\phi -Vt)
\biggr ],
\end{eqnarray}
which is the soliton solution of amplitude $\eta _0 $. Small oscillation occur when
$\alpha _3 \rightarrow \alpha _2 $ and $m \rightarrow 0 , T\rightarrow \pi
/2$.
Consequently, the system has two limiting solutions, a periodic and 
a localized traveling profile, which deform one into the other,
by  the initial conditions and the velocity parameter $V$. A figure showing the
deformation from the
$l=5$ cnoidal mode towards a soliton is shown in Figs. 2.

The cnoidal solution eq.(3) depends on the  parameters $\alpha _i$
subjected to the
volume conservation and the  periodicity condition of the solution (for the final soliton state this
condition should be taken as a  quasi-periodicity realised by the rapidly decreasing profile. This a
problem of the basic model, [10]). The periodicity restriction reads
\begin{eqnarray}
K\biggl ( \sqrt{{\alpha _3 -\alpha _2 }\over {\alpha _3 -\alpha _1 }}\biggr ) =
{{\pi}\over {n}}\sqrt{\alpha _3 -\alpha _1 }, \ \ \ n=1,2,\dots,  2\sqrt{\alpha _3 -\alpha _1 } .
\end{eqnarray}
Hence, a single free parameter remains, which
can be taken either one out of the three $\alpha
$'s, $V$ or $\eta _0 $.
Equatorial  cross-sections of the drop are shown in Fig. 2b for the
cnoidal solution at several values of the parameter $\eta _0$. 
All explicite calculations are presented in detail in [10].

In Fig. 3 we present the total energy plotted versus the parameters $\alpha _1 , \alpha _2$ for constant
volume. From the small oscillation limit ($\alpha _2 \simeq 3$ in the figure) towards the solitary wave
limit ($\alpha _2 =1$ in the figure) the energy increases and has a valley for $\alpha _1 \simeq 0.1$
and  $\alpha_2 \in (1.2, 1.75)$ (close to the $l=2$ mode). 
In order to introduce more realistic results, the total hydrodynamic energy is plotted versus 
$\alpha _1 ,\alpha _2$ for constant volume, too  but we marked those special solutions 
fulfilling the periodicity condition. In Fig. 4 we present the total energy valley, from the small
oscillations limit  towards the solitary wave limit. We notice that 
the energy constantly increases but around  $\alpha _2 \in (1.2, 1.75)$ (close to the
linear $l=2$ mode) it has a valley providing some stability for solitary
solution (also called roton [13]).

\section{The three-dimensional nonlinear  model}

In the following we introduce a sort of generalized KdV equation for fluids.
We consider the three-dimensional irrotational flow of an ideal incompressible fluid layer
in a semi-finte rectangular channel subjected to uniform vertical gravitation ($g$ in $z$ direction)
and to surface pressure [12]. The
depth of the layer, when the fluid is at rest is $z=h$. Boundary conditions at the finite spaced walls
consist in annilation of the  normal velocity component, i.e. on the bottom of the layer ($z=0$) and on 
the walls $x=x_{0}\pm L/2$ of the channel of width $L$. The following results remain valid if the
walls expand arbitrary, e.g. $L\rightarrow \infty$, and the flow is free. We choose for the
potential of the velocities the form
\begin{eqnarray}
\Phi =\sum_{k\ge 0}\alpha _k (t)
\cos {{k\pi (x-x_0 )} \over L}\cosh {{\sqrt{2}k\pi (y-y_0 )} \over L}\cos {{k\pi z} \over L},
\end{eqnarray}
where $\alpha _k (t)$ are arbitrary functions of time and $L$ is a free parameter.
Eq.(7) fulfils $\triangle \Phi =0$ and the above boundary conditions at the walls. However there is
another boundary condition at the free surface of the fluid [7]
\begin{eqnarray}
( \Phi _z -\eta _t -\eta _x \Phi _x )_{z=h+\eta }=0,
\end{eqnarray}
where $\eta (x,t)$ describes the shape of the free surface. By introducing the
function
\begin{eqnarray}
f(x,t)=\sum_{k=0}^{\infty }{{\alpha _k (t)k\pi } \over L}
\biggl ( \sin {{k\pi (x-x_0 )} \over L}\cosh {{\sqrt{2}k\pi (y-y_0 )} \over L}\biggr ) , 
\end{eqnarray}
the velocity field on the free surface can be written
\begin{eqnarray}
\Phi _x |_{z=h+\eta } =- \cosh(z\partial _{x})f , \nonumber \\ 
-\Phi _z |_{z=h+\eta }=-\sinh(z\partial _{x})f .
\end{eqnarray}
Eqs (10) do not depend on $L$ and the case $L \rightarrow \infty $ of unbounded channels and free
travelling profiles remains equaly valid. Since the unique force field in the problem is
potential, the dynamics is described by the Bernoulli equation, which, at the free surface,
reads
\begin{eqnarray}
\Phi _{xt}+\Phi _{x}\Phi _{xx}+\Phi _{z}\Phi _{xz}+g{\eta }_{x}+{1 \over {\rho }}P_{x}=0.
\end{eqnarray}
Here $P$ is the surface pressure obtained by equating
$P$'s first variation with the local mean curvature of the surface, under the restriction of the
volume conservation
\begin{eqnarray}
P\bigg |_{z=h+\eta }={{\sigma {\eta }_{xx}} \over {{(1+{\eta }_{x}^{2})}^{3/2}}},
\end{eqnarray}
and $\sigma $ is the surface pressure coefficient. The pressure in  eq.(12) approaches $-\sigma {\eta
}_{xx}$, for small enough relative amplitude of the deformation
$\eta /h$.  In order to solve the system of the two partial differential equations (8,11) with respect
to the unknown functions $f(x,t)$  and $\eta (x,t)$, we consider the approximation of small
perturbations of the surface compared to  the depth, $a=max|{\eta }^{(k)}(x,t)|<<h$,  where $k=0,...,3$
are orders of differentiation. Inspired by [12] and using a sort of perturbation  technique in $a/h$, we
obtain from eqs.(6-11) the  generalised KdV equation
\begin{eqnarray}
{{\eta }_{t}+{{c_{0}}\over h}\sin(h\partial ){\eta } +  {{c_{0}}\over h}({\eta }_{x}\cosh(h\partial )
{\eta }} +\eta \cosh (h\partial ){\eta }_{x})0. 
\end{eqnarray}
If we approximate $\sin (h\partial )\simeq h\partial -{1 \over  6}(h\partial )^{3}$, $\cosh (h\partial
)\simeq 1-{1 \over 2} (h\partial )^{2}$, we obtain,  from eq.(9), the polynomial differential equation:
\begin{eqnarray}
a{\tilde {\eta }}_{t}+2c_{0}{\epsilon }^{2}h{\tilde {\eta }}
{\tilde {\eta }}_{x}+c_{0}\epsilon h{\tilde {\eta }}_{x}
-c_{0} \epsilon {{h^{3}} \over 6}{\tilde {\eta }}_{xxx}-
{{c_{0}{\epsilon }^{2}h^{3}} \over 2}\biggl (
{\tilde {\eta }}_{x}{\tilde {\eta }}_{xx}+{\tilde {\eta }}{\tilde 
{\eta }}_{xxx}\biggr ) & = & 0, \hfill
\end{eqnarray}
where $\epsilon ={a \over h}$. The first four terms in eq.(20)  correspond  to the zero order
approximation terms in $f$, obtained  from the boundary  condition at the free surface,  i.e. the
traditional way of obtaining the KdV equation in shallow channels.

In order to find an exact solution for  eq.(12) we can  write it in the form:
$$
Ah{u }_{X}(X)+{{u (X+h)-u (X-h)} \over {2i}}
+{u }_{X}(X){{u (X+h)+u (X-h)} \over 2} 
$$
\begin{eqnarray}
+u (X)
{{{u }_{X}(X+h)+{u }_{X}(X-h)
} \over {2}}=0, 
\end{eqnarray}
where $X=x+Ac_0 t$ and $A$ is an arbitrary real constant.
We want to stress here that eq.(14) is a finite-difference differential equation, which is rather the
exception than the rule fir such systems. Hence, it may contain among its symmetries, the scaling
symmetry. Actualy, the first derivative of $u(X)$ is shown to be alinear combination of translated
versions of the original function. In this way, the theory of such equations can be related with the
wavelet, or other self-similarity systems, theory, [13]. In the following we study the solutions with a
rapid decreasing  at infinity and make a change of variable:
$v=e^{BX}$ for
$x \in (-\infty ,0)$ and $ v=e^{-BX}$ for $x \in (0, \infty )$, with  $B$ an arbitrary constant.
Writing $u(X)=-hA+f(v)$, and choosing for the solution the form
of a power series in $v$:
\begin{eqnarray}
f(v)=\sum_{n=0}^{\infty }a_{n}v^{n},
\end{eqnarray}
we obtain a nonlinear recurrsion relation for the
coefficients $a_{n}$:
$$
\biggl ( Ahk+{{\sin(Bhk)} \over {B}}\biggr ) a_{k} 
$$
\begin{eqnarray}
=-\sum_{n=1}^{k-1}n\biggl (\cosh\left( Bh(k-n)\right )+\cosh (Bh(k-1))\biggr )a_{n}a_{k-n}. 
\end{eqnarray}
With the coefficients given in eq.(16) the general solution $\eta $ can be written analyticaly.
In order to verify the consistency of this solution
we study a limiting case of the relation, by replacing 
$\sin$  and $\cosh$ expressions with their lowest nonvanishing terms in their power 
expansions
Thus, eq.(16) reduces to
\begin{eqnarray}
{\alpha }_{k}={6 \over {B^{2}h^{3}k(k^{2}-1)}}
\sum_{n=1}^{k-1}n{\alpha }_{n}{\alpha }_{k-n},
\end{eqnarray}
and
\begin{eqnarray}
{\alpha }_{k}=
\biggl ( {1 \over {2B^{2}h^{3}}}\biggr )^{k-1}k  \hfill
\end{eqnarray}
is the solution of the above recurrence relation.
In this approximation, the solution of eq.(12) reads
\begin{eqnarray}
\eta (X) & = & 2B^{2}h^{3}\sum_{k=1}^{\infty }k
\biggl ( -e^{-B|X|}
\biggr )^{k}={{B^{2}h^{3}} \over 2}{1 \over {\left(
\cosh(BX/2)\right)^{2}}},
\end{eqnarray}
which is just the single-soliton solution of the
KdV equation and it was indeed obtained by assuming $h$ small 
in the recurrence relation (16).
Hence, we have shown that the KdV equation describing the shallow liquids can be generalised for any
depths and lengths. This result may be the starting point to search for more interesting symmetries. 
It would be interesting to interpret the generalized-KdV eq.(12) as the Casimir element of a certain
algebra.

\section{Compacton and self-similar solutions}

Eq.(12) has a special character, namely contains both infinitesimal and finite difference operators.
This particularity relates it to another field of nonlinear systems, that is scaling functions and
wavelet basis, functions or distributions with compact support and self-similarity properties. In the
following we investigate a particular case of eq.(12), that is when
$h\ll
\eta$,
$h\ll \delta$, where $\delta $ is the half-width of the solution, if this has bounded or compact
support. In this approximations, from eq.(12) we keep only the terms
\begin{eqnarray}
{1 \over {c_0 }}\eta _t +\eta _x +{1 \over h} \eta \eta _x -{{h}\over {2}}\eta _x \eta _{xx}
+{1 \over h} \eta \eta _x -{h \over 2} \eta \eta _{xxx} +{\cal O}_{3}\simeq 0.
\end{eqnarray}
This equation is related to another intergable system, namely the K(2,2) equation, investigated in
[11]
\begin{eqnarray}
\eta _t +(\eta ^2 )_x +(\eta ^2)_{xxx}=0.
\end{eqnarray}
The main property of the K(2,2) equation is the equal occurence of non-linearity,
dispersion and the existence of a Lagrangian and Hamiltonian system associated with it. The
special solutions of this equation are the compactons
\begin{eqnarray}
\eta _{c}={{4 \eta _{0}}\over {3}}\cos ^2 \biggl ( {{x-\eta _{0}t}\over {4}} \biggr )  , \ \ \
|x-\eta _{0} t|\geq 2\pi ,
\end{eqnarray}
and $\eta _c =0$ otherwise.
This special solutions have compact support and special
properties concerning the scattering between different such solutions.
As the authors comment in [11], the robustness of these solutions makes it clear that a new mechanism is
underlying this system. In this respect, we would like to add that, taking into account eq.(12),
this new mechanism might be related to selfsimilarity and multiscale properties of nonlinear systems.

\vskip 1cm
\section{Conclusions}
\vskip 1cm

In the present paper we introduced a non-linear hydrodynamic model describing new modes of  motion of
the free surface of a liquid. The total energy of this nonlinear liquid drop model, subject to
non-linear boundary conditions at the free surface and the inner surface of the fluid layer,
gives the Hamiltonian of the Korteweg de Vries equation. We have studied the stability of the cnoidal
wave and solitary wave solutions, from the point of view of minima of this Hamiltonian.

The non-linear terms yield rotating steady-state solutions that are cnoidal waves on the surface
of the drop, covering continuously the range from small harmonic oscillations, to anharmonic 
oscillations, and up to solitary waves.
The initial one-dimensional model [10] was extend to a three-dimensional model. A kind of new
generalized KdV equation, together with some of its analytical solutions have been presented.
We also found a connection between the obtained generalized KdV equation, and another one (i.e. K(2,2)),
in a certain approximation. In this case, compacton solutions have been found and new symmetries (e.g.
self-similarity) were put into evidence.

The analytic solutions of the  non-linear
model presented in this paper,  make  possible the study of clusterization
as well as to explain or predict the existence of new strongly deformed shapes, or new patterns having
compact support or finite wavelength. The model applies not only in fluid and rheology theories, but may
provide insight into similar processes occurring in other fields and at other scales, such as the
behavior of superdeformed nuclei, supernova, preformation of cluster in hydrodynamic models (metallic,
molecular, nuclear), the fission of liquid drops (nuclear physics), inertial fusion, etc.

\vskip 1cm
Supported  by the U.S. National Science Foundation through a regular grant, No. 9603006, and a 
Cooperative Agreement, No. EPS-9550481, that includes a matching component from the Louisiana
Board of Regents Support Fund.  One of the authors (A.L.) would like to thank Peter Herczeg from the
T5 Division at Los Alamos National Laboratory, and at the Center for Nonlinear Studies at Los
Alamos for hospitality.

\vfill
\eject

\centerline{FIGURE CAPTIONS}
\vskip 1cm
Fig. 1

The order of smallness of four typical terms depending on $\phi $ and occuring
in the Hamiltonian, eqs.(1,2).  Order zero holds for $\eta ^2$, order 1 for $\eta
_{\phi }^{2}$, order 2 for $\eta ^3, $ and order 3 for $\eta \eta _{\phi }^{2}$.

\vskip 1cm
Figs. 2

2a.
 
The transition of the cnoidal solution, from a $l=5$ mode to the soliton limit:  shape of the
cross-section for $\theta =\pi /2$ function as a function of $\alpha _2 $ with $\alpha _{1,3}$ fixed by
the volume conservation and periodicity conditions.

2b.

Cnoidal solutions (cross-sections of $\Sigma 1$ for $\theta =\pi /2$) subject to the
volume conservation constraint. Results for the $l=6$ mode to the $l=2$ mode and a soliton are
shown. The corresponding linear modes, i.e. spherical harmonics, are superimposed on the
non-linear solutions.

2c.

Pictorial view of a soliton deformation of a drop, on the top of the original undeformed sphere.
The supporting sphere for the soliton has smaller radius  because of the volume conservation.

\vskip 1cm

\vskip 1cm
Fig. 3

The energy plotted versus $\alpha _1 , \alpha _2$ for constant volume. From the
small oscillation limit ($\alpha _2 \simeq 3$) towards the solitary wave limit ($\alpha _2
=1$) the energy increases and has a valley for $\alpha _1 \simeq 0.1$ and 
$\alpha_2 \in (1.2, 1.75)$ (close to the $l=2$ mode). 

\vskip 1cm
Fig. 4

The total energy plotted versus $\alpha _1 ,\alpha _2$ for
constant volume (small circles). Larger circles indicate the patterns
fulfilling the periodicity condition. From the small oscillations limit
($\alpha _2 \simeq 3$) towards the solitary wave limit ($\alpha _2 =1$)
the energy increases but for $\alpha _2 \in (1.2, 1.75)$ (close to
$l =2$ mode) it has a valley..


\begin{thebibliography}{99}

\bibitem{1}
H-L. Lu and R. E. Apfel, {\it {J. Fluid Mech.}} {\bf {222}} 351 (1991);
T. Shi and R. E. Apfel, {\it {Phys. Fluids}} {\bf {7}} 1545 (1995);
Y. Tian, R. G. Holt and R. E. Apfel, {\it {Phys. Fluids}} {\bf {7}} 2938 (1995);
W. T. Shi, R. E. Apfel and R. G. Holt, {\it {Phys. Fluids}} {\bf {7}} 2601 (1995).

\bibitem{2}
R. Natarajan, R. A. Brown, {\it {J. Fluid Mech.}} {\bf {183}} 95 (1987); {\it {Phys. Fluids}}
{\bf {29}} 2788 (1986).

\bibitem{3}
J. A. Tsamopoulos and R. A. Brown, {\it {J. Fluid Mech.}} {\bf {127}} 519 (1983).

\bibitem{4}
E. H. Trinh, R. G. Holt and D. B. Thiessen, {\it {Phys. Fluids}} {\bf {8}} 43 (1995);
P. L. Marston and S. G. Goosby, {\it {Phys. Fluids}} {\bf {28}} 1233 (1985);
E. Trinh and T. G. Wang, {\it {J. Fluid Mech.}} {\bf {122}} 315 (1982).

\bibitem{5}
R. E. Apfel {\it {et al}}, {\it {Phys. Rev. Lett.}} {\bf {78}} 1912 (1997)

\bibitem{6}
E. Trinh and T. G. Wang, {\it {J. Fluid Mech.}} {\bf {122}} (1982) 315

\bibitem{7} 
G. L. Lamb, {\it Elements of Soliton Theory} (John Wiley \& Sons, New York, 1980);  C. Rebbi
and G. Soliani, {\it Solitons and Particles} (World Scientific, Singapore, 1984);  S.
Novikov, S. V. Manakov, {\it Theory of Solitons: The Inverse Scattering Method} (Consultants
Bureau, New York, 1984);  R. K. Bullough and P. J. Caudrey, Eds., {\it {Solitons}} (Topics in
Current Physics, Springer-Verlag, Berlin, 1980).

\bibitem{8}
A. Ustinov, {\it {Solitons in Josephson Junctions}}, 
{\it {non-linear Waves and Solitons in Physical Systems}}
(May 12-16, 1997, CNLS, Los Alamos, NM) to appear in {\it {Physica D}}.

\bibitem{9}
Y. G. Kevrekidis, {\it{Catalytic Pattern on Microdesigned Domains}},
{\it {non-linear Waves and Solitons in Physical Systems}}
(May 12-16, 1997, CNLS, Los Alamos, NM) to appear in {\it {Physica D}}.

\bibitem{10}
A. Ludu, A. Sandulescu and W. Greiner, {\it {Int. J. Modern Phys. E}} {\bf 1} 169 (1992) ; 
{\bf 2} 4 (1993) 855;  {\it J. Phys. G: Nucl. Part. Phys.} {\bf 21} 1715 (1995); {\it J. Phys. G: Nucl.
Part. Phys.} {\bf 23} 343 (1997).

\bibitem{11}
P. Rosenau and J. M. Hyman, {\it {Phys. Rev. Let.}} {\bf {70}} 564 (1993);
F. Cooper, J. M. Hyman and A. Khane, to be published.

\bibitem{12}
A. Ludu and W. Greiner, {\it {Found. Phys.}} {\bf {26}} 665 (1996).

\bibitem{13}
A. Ludu and J. P. Draayer, in preparation; Internal seminar CNLS and T8, Los Alamos, May, 1996.

\bibitem{14}
R. Abraham and J. E. Marsden, {\it Foundations of Mechanics}
(The Benjamin/Cummings Publishing Company, Inc., Reading, Massachusetts,
1978).

\end{thebibliography}
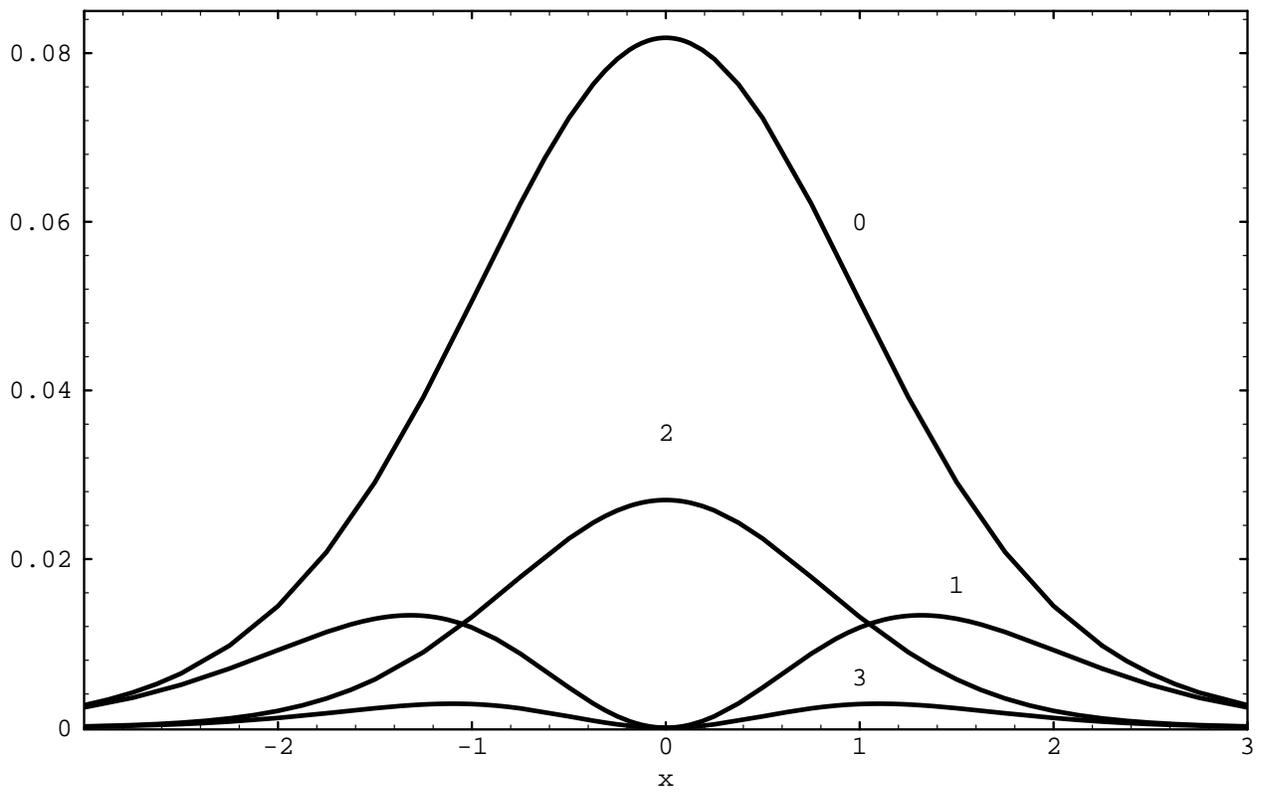
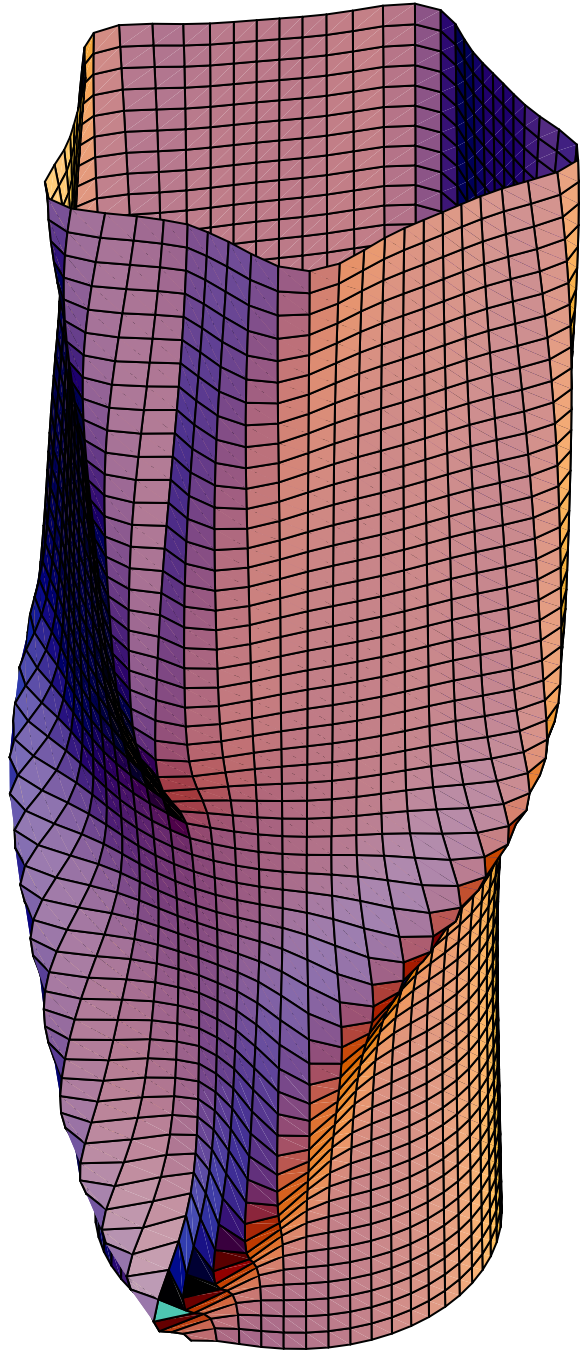
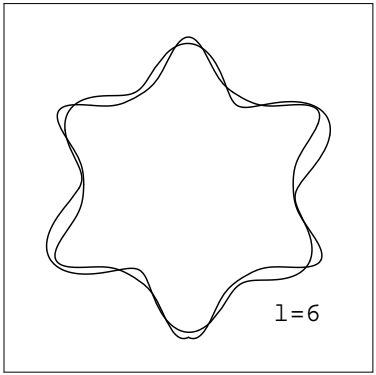
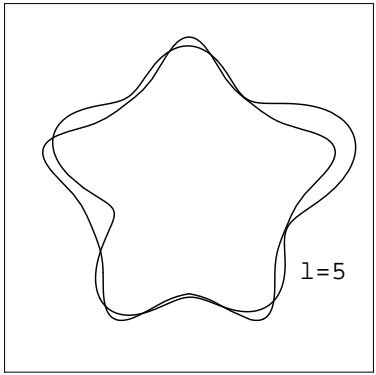
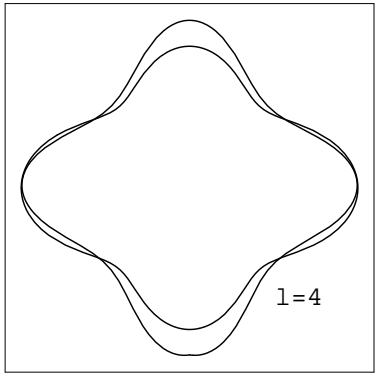
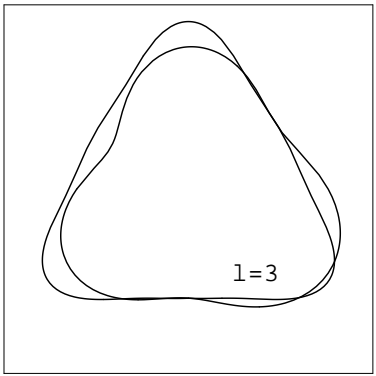
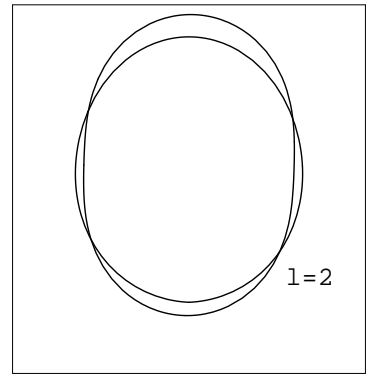
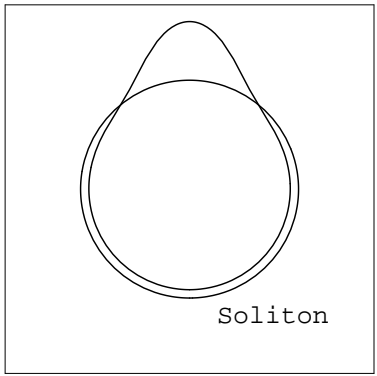
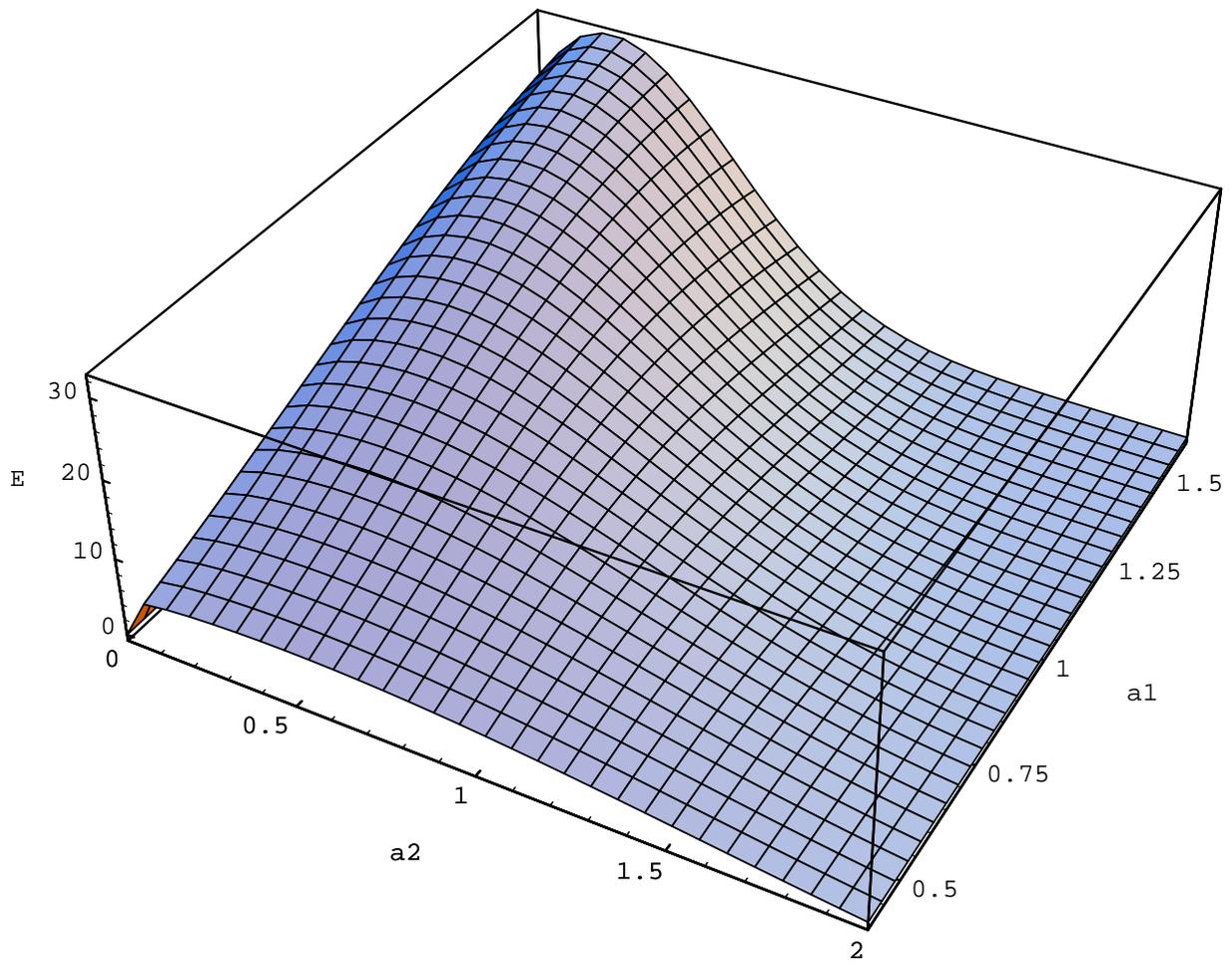
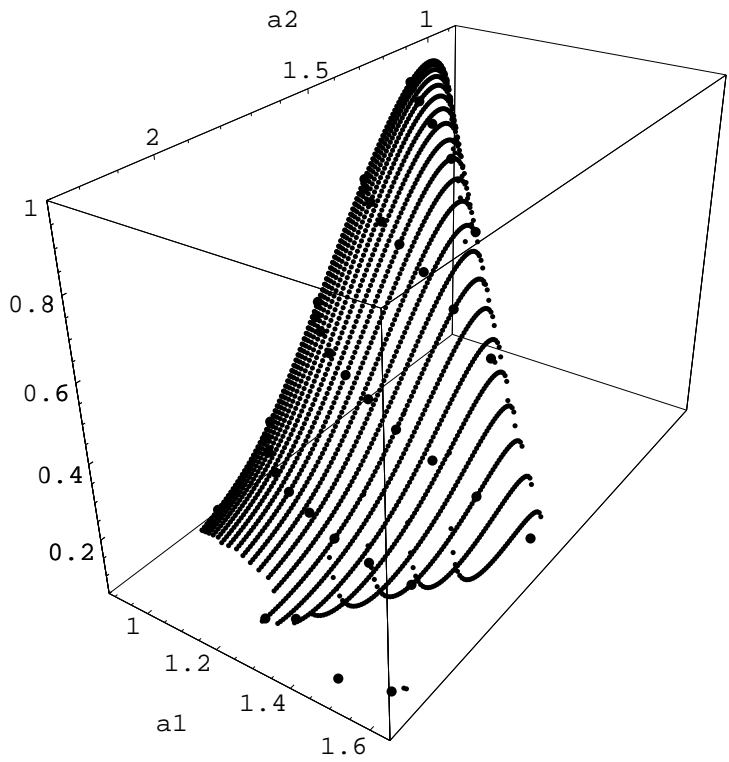
\end{document}